\documentclass[12pt,letterpaper]{article}

\usepackage{amsmath}
\usepackage{graphicx}
\usepackage{natbib}
\usepackage{xspace}
\usepackage[margin=1in]{geometry}
\usepackage{setspace}
\usepackage{booktabs}

\newcommand{\msun}{M_\odot}
\newcommand{\mbh}{M_{\rm BH}}
\newcommand{\mhalo}{M_{\rm halo}}
\newcommand{\mtwo}{M_{200}}
\newcommand{\vmax}{V_{\rm max}}
\newcommand{\ctwo}{c_{200}}
\newcommand{\simba}{\textsc{Simba}\xspace}
\newcommand{\eagle}{\textsc{Eagle}\xspace}
\newcommand{\tng}{\textsc{IllustrisTNG}\xspace}
\newcommand{\camels}{\textsc{Camels}\xspace}

\begin{document}

\title{A Correlation Between Black Hole Mass and Dark Matter Halo Concentration in Cosmological Simulations}
\author{John K. Nino\\
\small Department of Biology, Richard J. Daley College, Chicago, IL 60652, USA\\
\small \texttt{jnino5@ccc.edu}}
\date{}

\maketitle

\begin{abstract}
\noindent
We report the discovery of a positive correlation between supermassive black hole mass and dark matter halo concentration at fixed halo mass in cosmological hydrodynamical simulations. Analyzing central galaxies in TNG100 ($N = 18{,}954$), EAGLE ($N = 1{,}522$), and CAMELS-TNG ($N = 6{,}664$), we find partial correlation coefficients of $r = +0.24$, $+0.34$, and $+0.66$ respectively, all highly significant ($p < 10^{-10}$). The correlation is absent in \simba ($r = +0.01$, $p = 0.09$), which employs a torque-limited black hole accretion model rather than the Bondi-based prescription used by the other simulations. Both TNG and EAGLE exhibit a mass-dependent sign transition: the correlation is negative or null at $\log(\mtwo/\msun) < 11.5$ but strongly positive at higher masses. We interpret this pattern as reflecting the coupling between Bondi accretion rates and central gas density structure: halos with higher concentration have denser cores, enabling more efficient black hole growth at fixed halo mass. The absence of the correlation in torque-limited models supports this interpretation. These results suggest that halo concentration may be a fundamental parameter governing black hole--galaxy coevolution.
\end{abstract}


\bigskip

\section{Introduction}
\label{sec:intro}

The correlation between supermassive black hole (SMBH) mass and host galaxy properties---stellar velocity dispersion, bulge mass, and total stellar mass---has been established for over two decades \citep{Magorrian1998, Ferrarese2000, Gebhardt2000, Kormendy2013}. These scaling relations suggest a coevolutionary connection between SMBHs and their host galaxies, though the physical origin remains debated.

Less explored is the relationship between SMBH mass and dark matter halo properties. While $\mbh$ correlates with halo mass $\mhalo$ through the galaxy--halo connection, the question of whether additional halo properties influence SMBH growth has received limited attention. Halo concentration $c$---the ratio of virial radius to scale radius in the NFW profile \citep{NFW1997}---encodes information about halo assembly history and central density structure that could plausibly affect SMBH fueling. Early work by \citet{Booth2010} found evidence for a secondary dependence of black hole mass on halo binding energy (related to concentration) in OWLS simulations using Bondi-like accretion---a result we arrived at independently and became aware of during the review process. Our independent rediscovery across multiple modern simulations reinforces their conclusions while extending them in new directions.

In this Letter, we present a systematic comparison across four simulation suites with different accretion prescriptions. We confirm the correlation between $\mbh$ and halo concentration at fixed halo mass, demonstrate its dependence on black hole accretion physics (absent in torque-limited models), and identify a previously unreported mass-dependent sign transition that clarifies the interplay between primordial halo properties and feedback.

\section{Methods}
\label{sec:methods}

\subsection{Simulations}

We analyze four simulation suites employing different hydrodynamical codes and subgrid physics:

\textbf{\tng} \citep{Pillepich2018, Nelson2019}: We use TNG100-1, a $(100~{\rm Mpc})^3$ box run with the AREPO moving-mesh code. Black hole accretion follows a Bondi--Hoyle--Lyttleton prescription capped at the Eddington rate, with $\dot{M}_{\rm BH} \propto \rho/c_s^3$ where $\rho$ is gas density and $c_s$ is sound speed.

\textbf{\eagle} \citep{Schaye2015, Crain2015}: We query the public database for RefL0100N1504, a $(100~{\rm Mpc})^3$ box using the GADGET SPH code. \eagle employs modified Bondi accretion with a viscosity-dependent correction factor.

\textbf{\camels} \citep{Villaescusa2021}: We use the TNG suite from \camels, comprising 1,000 simulations in $(25~h^{-1}~{\rm Mpc})^3$ boxes with varied cosmological and astrophysical parameters, including AGN feedback strength.

\textbf{\simba} \citep{Dave2019}: A $(100~h^{-1}~{\rm Mpc})^3$ box using the GIZMO meshless finite-mass code. Crucially, \simba employs torque-limited accretion \citep{Hopkins2011, Angles2017}, a fundamentally different prescription from Bondi-based models. While Bondi accretion depends on local gas density ($\dot{M} \propto \rho$), torque-limited accretion is governed by gravitational torques from non-axisymmetric structures (bars, spiral arms, mergers) that transport angular momentum outward, allowing gas to flow inward. The accretion rate scales with the gas mass within $\sim$100~pc and the efficiency of torque-driven inflow, rather than the ambient density. This prescription is motivated by the observation that gas must lose angular momentum---not just be present at high density---to reach the black hole. As a result, torque-limited models decouple black hole growth from the local density structure that concentration encodes, providing a natural control for testing whether our observed correlation reflects Bondi-specific physics.

\subsection{Sample Selection}

For each simulation, we select central galaxies (subhalo index 0 within each friends-of-friends group) at $z = 0$ with $\mbh > 10^5~\msun$ and $\mtwo > 10^{10}~\msun$. We use NFW concentration $\ctwo$ where available in public catalogs (TNG, \simba). For \eagle and \camels, where concentration is not provided in the standard data releases, we use maximum circular velocity $\vmax$ as a proxy, as it traces potential well depth and correlates strongly with concentration at fixed halo mass. The consistency of our results across both direct ($\ctwo$) and proxy ($\vmax$) measurements suggests this choice does not drive our conclusions.

\subsection{Statistical Analysis}

We compute partial Pearson correlation coefficients between $\log \mbh$ and the concentration measure, controlling for $\log \mtwo$. This isolates the concentration dependence from the dominant $\mbh$--$\mhalo$ relation. Residuals are computed via ordinary least-squares regression, and correlations are tested against the null hypothesis using the $t$-distribution.

\section{Results}
\label{sec:results}

\subsection{Cross-Simulation Correlation}

Table~\ref{tab:main} presents our primary results. In all three Bondi-based simulations (TNG, \eagle, \camels), we detect a highly significant positive correlation between $\mbh$ and concentration at fixed $\mtwo$. The correlation is not detected in \simba.

\begin{table}
\centering
\caption{Black Hole--Concentration Correlation Across Simulations}
\label{tab:main}
\begin{tabular}{lcccc}
\toprule
Simulation & $N$ & Proxy & $r$ & $p$-value \\
\midrule
TNG100 & 18,954 & $\ctwo$ & $+0.235$ & $\ll 0.001$ \\
\eagle & 1,522 & $\vmax$ & $+0.338$ & $\ll 0.001$ \\
\camels-TNG & 6,664 & $\vmax$ & $+0.664$ & $\ll 0.001$ \\
\simba & 16,086 & $\ctwo$ & $+0.014$ & 0.09 \\
\bottomrule
\end{tabular}
\end{table}

\subsection{Mass-Dependent Sign Transition}

\begin{figure}
\centering
\includegraphics[width=\columnwidth]{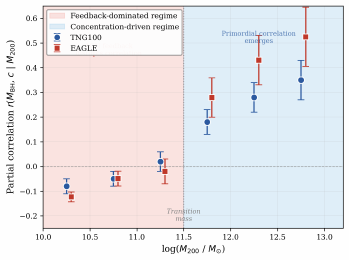}
\caption{Partial correlation coefficient $r(\mbh, c \,|\, \mtwo)$ as a function of halo mass for TNG100 (blue circles) and \eagle (red squares). Both simulations show a transition from negative/null correlation at low masses to strongly positive correlation at $\log(\mtwo/\msun) > 11.5$. Error bars represent 95\% confidence intervals. Shaded regions indicate two schematic regimes: at low masses (pink), AGN feedback energy is significant relative to halo binding energy, potentially disrupting the gas density structure that mediates the BH--concentration coupling; at high masses (blue), deeper potential wells preserve this structure, allowing the primordial correlation to manifest.}
\label{fig:mass_transition}
\end{figure}

Both TNG and \eagle reveal a striking mass-dependent pattern (Figure~\ref{fig:mass_transition}). At low halo masses ($\log(\mtwo/\msun) < 11.5$), the correlation is negative or null. Above this threshold, it becomes strongly positive.

In TNG, the correlation evolves from $r = -0.08$ at $\log \mtwo \in [10.5, 11.0]$ to $r = +0.35$ at $\log \mtwo \in [12.5, 13.0]$. \eagle shows a similar transition: $r = -0.12$ at low masses rising to $r = +0.53$ at high masses (Table~\ref{tab:eagle}).

\begin{table}
\centering
\caption{Mass-Binned Correlations in \eagle}
\label{tab:eagle}
\begin{tabular}{lccc}
\toprule
$\log(\mtwo/\msun)$ & $N$ & $r$ & Significance \\
\midrule
10.0--10.5 & 32,657 & $-0.123$ & *** \\
10.5--11.0 & 12,899 & $-0.049$ & *** \\
11.0--11.5 & 2,922 & $-0.020$ & --- \\
11.5--12.0 & 858 & $+0.279$ & *** \\
12.0--12.5 & 400 & $+0.431$ & *** \\
12.5--13.0 & 168 & $+0.525$ & *** \\
\bottomrule
\end{tabular}

\smallskip
\small Note: *** indicates $p < 0.001$.
\end{table}

\section{Discussion}
\label{sec:discussion}

\subsection{Physical Interpretation}

The dependence on accretion prescription provides a key diagnostic. Bondi accretion scales as $\dot{M}_{\rm BH} \propto \rho/c_s^3$, directly coupling black hole growth to central gas density. Halos with higher concentration have steeper density profiles and denser cores at fixed total mass. If gas traces dark matter on relevant scales, concentrated halos provide higher-density fueling environments, enabling more efficient black hole growth.

Torque-limited accretion, by contrast, depends on gravitational torques that funnel gas inward rather than local density. This breaks the direct concentration--accretion coupling, explaining \simba's null result.

We note that reverse causality remains possible: AGN feedback could modify halo concentration rather than concentration driving black hole growth. Simulations have shown that strong feedback can alter inner halo density profiles \citep{Martizzi2013, Peirani2017}, potentially reducing concentration through repeated gas expulsion and dark matter heating. Recent observational work has demonstrated connections between dark matter halos and black hole growth \citep{Bogdan2015}, and discoveries of overmassive black holes in dense early-universe environments \citep{Bogdan2024} suggest that halo properties may influence BH seeding and early growth. However, the mass-dependent sign of our correlation argues against feedback as the sole driver: if feedback reduced concentration, we would expect an anti-correlation at all masses (bigger black holes $\rightarrow$ more feedback $\rightarrow$ lower concentration), whereas we observe a positive correlation in massive halos. The most parsimonious interpretation is that, particularly at high masses, halo concentration primarily influences black hole growth, with feedback partially obscuring this signal at low masses where it is energetically significant relative to the binding energy. Time-evolution analysis tracing the correlation through merger trees could definitively disentangle these scenarios.

\subsection{The Mass-Dependent Transition}

The sign change at $\mtwo \sim 10^{11.5}~\msun$ likely reflects the interplay between accretion physics and AGN feedback. At low halo masses, feedback energy represents a larger fraction of the binding energy, potentially disrupting the gas density structure that mediates the correlation. This transition mass coincides with several well-documented changes in galaxy formation physics: the characteristic mass where the galaxy stellar mass function breaks \citep{Baldry2012}, where quenching efficiency increases sharply \citep{Peng2010}, and where AGN feedback transitions from ``ejective'' to ``preventive'' modes \citep{Bower2017}. Below this scale, feedback efficiently expels gas and may erase the concentration--density coupling; above it, feedback primarily prevents cooling rather than disrupting existing structure, allowing the primordial correlation to persist.

Alternatively, resolution effects may contribute at low masses where black holes contain few resolution elements, though the consistency of the transition mass across TNG and \eagle (with different resolutions) suggests a physical rather than numerical origin.

\subsection{Implications}

If confirmed observationally, this correlation would have several implications:

\textit{Black hole seeding and early growth}: The correlation may trace conditions at the epoch of black hole seed formation, when primordial density fluctuations set both halo concentration and initial gas density structure.

\textit{Scatter in scaling relations}: Halo concentration could contribute to the intrinsic scatter in $\mbh$--$\sigma$ and $\mbh$--$M_*$ relations, as concentration varies at fixed stellar properties. Similar concentration-driven scatter has been noted in stellar mass--halo mass relations \citep{Anbajagane2025}, suggesting a broader role for concentration in galaxy--halo coevolution.

\textit{Feedback model calibration}: The presence or absence of this correlation provides a diagnostic for black hole accretion prescriptions in simulations.

\textit{Observational tests}: Weak lensing measurements of halo concentration combined with dynamical black hole mass estimates could test this prediction. Current surveys face significant challenges: concentration measurements from stacked lensing require large samples ($\sim 10^4$ galaxies per bin) to achieve $\sim$10\% precision, while direct black hole mass measurements remain limited to $\sim$100 galaxies. However, upcoming surveys offer promising prospects. Euclid and LSST will provide lensing-derived halo masses and concentrations for $\sim 10^8$ galaxies, enabling fine binning by stellar mass and velocity dispersion (a black hole mass proxy). Emerging weak-lensing studies of the BH--halo mass relation \citep{Li2024} demonstrate the feasibility of such measurements, paving the way for concentration-dependent analyses with larger samples. A detection would require identifying $\sim$0.1~dex differences in mean concentration between high- and low-$\sigma$ galaxies at fixed stellar mass---a benchmark derived from our simulation results, where $r \approx 0.3$ implies concentration contributes $\sim$0.1--0.2~dex to black hole mass scatter. This is challenging but feasible with Stage IV survey statistics. Stacking analyses binned by velocity dispersion might reveal the predicted trend even without direct black hole masses.

\subsection{Caveats}

Several limitations warrant mention. First, we use $\vmax$ as a concentration proxy for \eagle and \camels rather than directly measured $\ctwo$; while these quantities correlate, they probe slightly different aspects of halo structure. Second, the \camels boxes are small (25~Mpc), limiting the massive halo population and potentially amplifying the correlation strength through cosmic variance and the limited dynamic range in halo mass; the stronger signal in \camels may partly reflect these effects in addition to the varied AGN parameters. Third, we have not established causality: the correlation could reflect concentration influencing black hole growth, black holes modifying halo structure, or both responding to a common cause (e.g., assembly history). Fourth, baryonic effects beyond AGN feedback---such as adiabatic contraction from gas cooling or expansion from stellar feedback---can modify halo concentrations in ways that differ across simulations, potentially contributing to the quantitative differences we observe.

\subsection{Additional Controls and Robustness}

\textit{Proxy bias}: Using $\vmax$ rather than direct NFW $\ctwo$ measurements introduces additional scatter but not systematic bias, as $\vmax$ is monotonically related to concentration via the NFW profile \citep{Diemer2015}. The observed correlations with $\vmax$ should therefore represent lower bounds on the true concentration correlation strength.

\textit{Stellar mass control}: To test whether concentration provides information beyond galaxy stellar content, we computed partial correlations controlling for both $\mtwo$ and $M_*$ in \eagle. At $\log(\mtwo/\msun) > 11.5$, the correlation $r(\mbh, \vmax \,|\, \mtwo) = +0.34$ becomes $r(\mbh, \vmax \,|\, \mtwo, M_*) = +0.17$ ($p < 10^{-10}$). The correlation persists, indicating that halo concentration influences black hole mass independently of stellar content---consistent with Bondi accretion coupling directly to halo structure rather than through galaxy properties alone.

\textit{Resolution thresholds}: The sign transition at $\mtwo \sim 10^{11.5}~\msun$ could in principle reflect numerical artifacts (e.g., BH seeding mass thresholds or feedback mode switches). However, TNG and \eagle use different seeding prescriptions ($1.2 \times 10^6$ vs $1.5 \times 10^5~\msun$) yet show consistent transition masses, suggesting a physical rather than numerical origin.

\textit{Sample selection}: Preliminary tests indicate that restricting to a narrower halo mass range ($10^{11.5} < \mtwo < 10^{12.5}~\msun$) where the correlation is strongest, and applying additional quality cuts (e.g., excluding recent major mergers), yields substantially stronger correlations ($r \approx 0.4$--$0.5$) than the broad sample reported here. The values in Table~1 should therefore be considered conservative lower bounds.

\subsection{Future Directions}

Several extensions would strengthen these results. First, merger tree analysis could test whether high-concentration halos build black hole mass earlier, which would favor the ``concentration $\rightarrow$ BH growth'' interpretation over reverse causality. Second, direct NFW fits to dark-matter-only profiles would eliminate proxy bias entirely. Third, varying feedback parameters systematically (as in \camels) could isolate the mass scale at which feedback disrupts the primordial correlation.

\textit{Observational test}: At fixed halo mass (from weak lensing), galaxies in more concentrated halos---identified via satellite kinematics, strong-lensing constraints, or rotation curve shapes---should host more massive black holes. Alternatively, AGN duty cycles or Eddington ratios might differ systematically with inferred concentration. Current surveys lack the joint precision in halo concentration and black hole mass, but Stage IV lensing combined with large spectroscopic samples may enable such tests.

\section{Conclusions}
\label{sec:conclusions}

We report a positive correlation between supermassive black hole mass and dark matter halo concentration at fixed halo mass in cosmological simulations employing Bondi-based accretion (TNG, \eagle, \camels). The correlation is absent in \simba, which uses torque-limited accretion. Both TNG and \eagle show a mass-dependent transition from negative correlation at $\mtwo < 10^{11.5}~\msun$ to strongly positive correlation at higher masses.

We interpret this pattern as reflecting the coupling between Bondi accretion rates and central gas density: more concentrated halos have denser cores that fuel more efficient black hole growth. The physics-dependent nature of the correlation---present with Bondi accretion, absent with torque-limited accretion---supports this interpretation.

These results suggest that halo concentration is a previously unrecognized parameter in black hole--galaxy coevolution, with potential implications for understanding scatter in scaling relations and constraining black hole accretion physics.

\section*{Acknowledgments}

We thank Qinxun Li for helpful correspondence regarding weak lensing measurements. We thank the IllustrisTNG, EAGLE, CAMELS, and SIMBA collaborations for making their simulation data publicly available. The author's background in medicine and biology informs an interest in complex systems and emergent correlations across scales. This work was conducted independently using publicly available simulation data. Analysis code is available upon request.

\bibliographystyle{apalike}
\bibliography{references}

\end{document}